\pdfoutput=1
\documentclass[conference]{IEEEtran}
\IEEEoverridecommandlockouts
\renewcommand\IEEEkeywordsname{Keywords}
\usepackage{cite}
\usepackage{amsmath,amssymb,amsfonts}
\usepackage{graphicx}
\usepackage{textcomp}
\usepackage{xcolor}
\usepackage{algorithm}
\usepackage{algpseudocode}
\usepackage[caption=false,font=footnotesize]{subfig}
\usepackage{booktabs}
\usepackage[hyphens]{url}
\usepackage[colorlinks=true,urlcolor=blue,linkcolor=black,citecolor=black]{hyperref}

\begin{document}

\title{SIDE: Sensor Impersonation Detection at the Edge via Sequence Prediction}

\author{
  \IEEEauthorblockN{Nahom M. Birhan}
  \IEEEauthorblockA{Department of Electrical and Computer Engineering\\
    Old Dominion University\\
    Virginia, USA\\
    nbirh002@odu.edu}
  \thanks{\textit{This work was conducted at Carnegie Mellon University in May 2022 as part of the author's M.S. research. The acronym \textsc{SIDE} was adopted for this 2026 arXiv release; the underlying experiments and results are from the original 2022 study. The original extended research report is available \href{https://drive.google.com/file/d/1Zh-IXxYftJmgB37RXKo7kX4sslCkgrov/view?usp=sharing}{online}.}}
}

\maketitle

\begin{abstract}
Some low-cost Internet of Things (IoT) sensor deployments lack device-level source authentication, leaving them vulnerable to impersonation or injected sensor readings. We present a lightweight approach to sensor impersonation detection in a small proof-of-concept study. We formulate detection as a sequence-prediction problem. A model with three LSTM layers and two fully connected layers is trained only on univariate temperature readings from a genuine sensor, and a window of readings is flagged when its mean absolute prediction error exceeds the mean genuine error by more than six standard deviations. The model is converted to TensorFlow Lite in three variants, non-quantized (554 KB), 16-bit weight quantized (298.5 KB), and 8-bit weight quantized (185 KB), targeting an Arduino Nano 33 BLE aggregator. The variants were evaluated with the TensorFlow Lite interpreter on the local server, since on-device execution of LSTM models was not yet supported by TensorFlow Lite for Microcontrollers at the time of the study. On a controlled testbed with the impostor sensor placed in a hotter outdoor location, the three variants reached detection accuracies of 99.980\%, 99.972\%, and 98.206\%, and each flagged the change point when a test sequence switched from genuine to impostor data. Quantization made the model smaller but slower in our measurements. The genuine and impostor distributions were well separated, so these results show detection of a controlled distribution shift and should not be read as evidence of general device authentication.

\end{abstract}


\section{Introduction}\label{sec:intro}
\noindent
IoT deployments stream large volumes of time-series data from sensors to the cloud. Edge computing moves computation closer to the data source and cuts bandwidth, latency, and the amount of raw data that leaves a site. Edge ML, or TinyML, pushes this one step further by running inference on the sensing devices themselves. We use it here for one security problem in sensor networks, detecting when a device impersonates a genuine sensor and injects its own readings.

Our motivation is the sensor deployment in which the link between a sensor and its aggregator carries no device-level source authentication. An attacker who can reach that link can replace the genuine reading stream with one of its own. We take a behavioral rather than cryptographic approach. A model trained on the genuine sensor's time series predicts what that sensor should report next, and a stream whose prediction error drifts away from the learned behavior gets flagged.

Sensor readings arrive at fixed intervals, which gives a univariate time series. Classical autoregressive and moving-average models can forecast such a series, but they describe a fixed, finite response to past samples. A recurrent neural network keeps a state that depends on the whole past sequence, and the Long Short-Term Memory (LSTM) cell~\cite{hochreiter1997} eases the vanishing-gradient problem that limits plain recurrent networks. We model the genuine sensor with LSTM layers followed by fully connected layers.

We kept the experiment small and controlled. An LM35 sensor indoors provides the genuine readings. A second LM35 sensor outdoors, in a hotter spot, plays the impostor. We train on genuine data only and convert the model to TensorFlow Lite for an Arduino Nano 33 BLE aggregator, in non-quantized, 16-bit, and 8-bit weight-quantized variants. For each variant we report model size, inference time, and detection accuracy, and we check whether the model flags the change point when a test sequence switches from genuine to impostor readings.

We set out to answer four questions.
\begin{enumerate}
    \item Can impersonation detection at the edge be posed as a sequence-prediction problem on genuine data alone?
    \item How should a threshold on the mean absolute error (MAE) be chosen to separate genuine from impostor windows?
    \item When a stream switches from genuine to impostor data, at what point does the model flag the transition?
    \item How do 16-bit and 8-bit weight quantization affect model size, inference time, and detection accuracy?
\end{enumerate}

One caveat applies throughout. The genuine and impostor sensors sat in different thermal environments, so what the experiment tests is detection of a controlled distribution shift tied to an impersonation scenario. It does not establish general device authentication, identification of a device independent of its measured environment, or robustness to an adversary who mimics the genuine distribution. On-device execution of the LSTM model was not completed; the three variants were evaluated with the TensorFlow Lite interpreter on the local server. Section~\ref{sec:discussion} spells out these limitations.

The rest of the paper follows the usual order. Section~\ref{sec:related} covers related work available at the time of the study, Section~\ref{sec:system} the testbed and threat model, Section~\ref{sec:method} the detection method, and Section~\ref{sec:data} the data. Section~\ref{sec:results} gives the results, Section~\ref{sec:discussion} discusses them and their limitations, and Section~\ref{sec:conclusion} concludes.

\section{Related Work}\label{sec:related}
\noindent
What follows is the literature we reviewed when doing the work. It covers the edge-versus-cloud trade-off, machine learning for IoT security, time-series modeling, and anomaly detection.

\subsection{Edge and Cloud}
Maamar et al.~\cite{cvse} conclude that the requirements of a given IoT application determine whether a cloud-only, edge-only, or combined deployment is appropriate. Shi et al.~\cite{edge1} argue that processing near the source saves bandwidth and reduces latency, energy, and cost. Nikolaou et al.~\cite{cost} compare edge and cloud deployments using a total-cost-of-ownership model in a wireless denial-of-service case study and report a 2.13-fold advantage for the edge. Silva et al.~\cite{stream} find that combining edge and cloud allows time-sensitive stream processing at the edge while less time-critical work runs in the cloud.

\subsection{Sensor Network Security and Machine Learning}
Chan et al.~\cite{sia} study secure information aggregation in sensor networks when a node is compromised. Their protocol considers a single aggregator and assumes an authenticated broadcast channel from the base station. Ullah et al.~\cite{imac} propose an implicit message authentication code for authenticating messages from IoT devices. Several surveys and studies argue that machine learning can improve IoT and edge security~\cite{useMLinCS,edge2,useML,useML1}. Wang et al.~\cite{edge2} evaluate the feasibility of machine learning for security at the IoT edge and find that a small deep neural network outperformed logistic regression in their TinyML setting. Efficient on-device inference depends on optimizations such as pruning, weight clustering, and quantization applied to trained networks~\cite{edge3}.

\subsection{Time-Series Modeling and Anomaly Detection}
Classical statistical models such as ARMA and ARIMA have long been used for time-series forecasting~\cite{ts1}. Lehna et al.~\cite{ts1} compare time-series and neural network models and suggest that combinations of the two can improve forecasts. LSTM networks, including bidirectional variants, have been applied to traffic forecasting~\cite{ts2,ts3}, and a recurrent neuro-fuzzy model has been proposed for time-series anomaly detection~\cite{ts4}. Closer to this work, Yu et al.~\cite{ay1} detect traffic anomalies in wireless sensor networks with an ARIMA model, Salem et al.~\cite{ay2} use Markov models for anomaly detection in wireless body area networks, Anagnostou et al.~\cite{ay3} detect deviations from normal power-system operation with an observer-based method, and Wu et al.~\cite{ay4} detect video anomalies with pretrained convolutional networks.

When we did this work we found little experimental work that trains a model on a genuine sensor's time series and then runs the detector on a microcontroller-class device to flag an impersonating source. The studies above deal with anomaly detection in other settings or with the edge-versus-cloud trade-off in general. Our contribution is a small, end-to-end experiment of that kind, including the effect of weight quantization on the deployed detector.

\section{System Setup and Threat Model}\label{sec:system}
\begin{figure}[!t]
    \centering
    \includegraphics[width=\columnwidth]{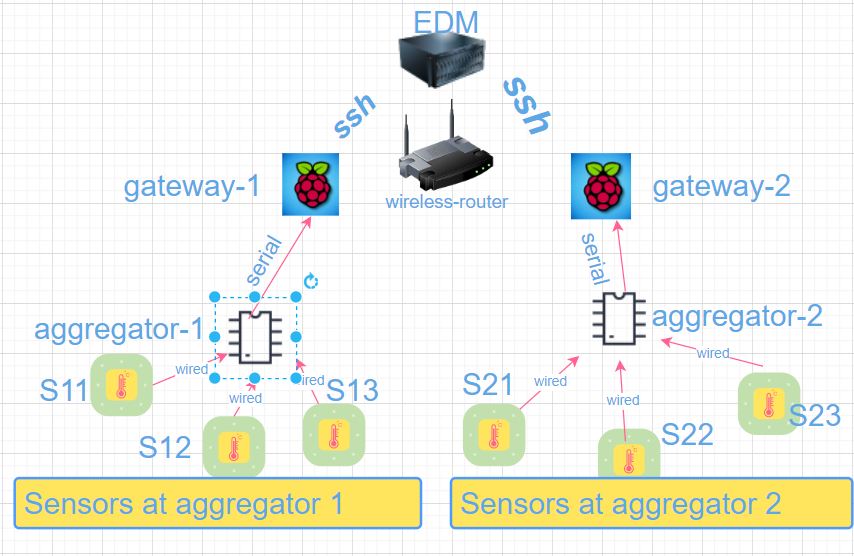}
    \caption{Testbed used for the experiments. Genuine data were collected through aggregator-1 (left) and impostor-condition data through aggregator-2 (right). Sensor-to-aggregator and aggregator-to-gateway links are wired; gateways reach the EDM over SSH through a wireless router.}
    \label{fig: Actual}
\end{figure}

\begin{figure}[!t]
    \centering
    \includegraphics[width=0.6\columnwidth]{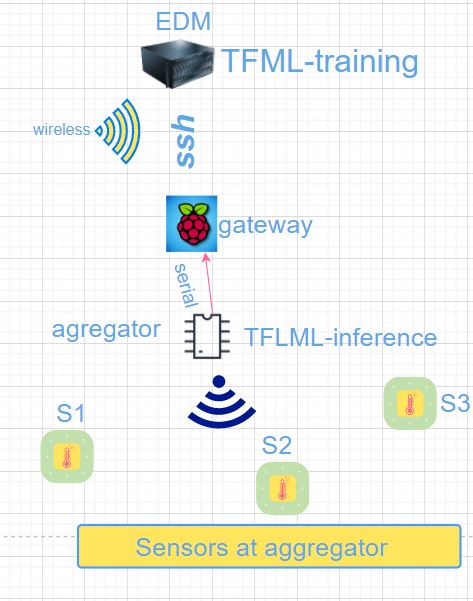}
    \caption{Assumed production configuration with wireless links between sensors and the aggregator. This configuration was not used in the experiments.}
    \label{fig: Ideal}
\end{figure}

\noindent
The testbed in Fig.~\ref{fig: Actual} has four kinds of components, an Edge Device Monitor (EDM), gateways, aggregators, and sensors. Fig.~\ref{fig: Ideal} shows the fully wireless configuration we had in mind for a production deployment. The experiments used the wired configuration of Fig.~\ref{fig: Actual}, since the LM35 sensors have no wireless interface.

\subsection{Components}
\textbf{Edge Device Monitor.} A personal computer acts as the EDM. It trains the TensorFlow model, converts it to TensorFlow Lite, and displays communication status, detection notifications, and the model's execution time and memory usage. The EDM communicates with the gateways over SSH and does not need an internet connection.

\textbf{Gateway.} Each gateway is a Raspberry Pi 4 Model B. It relays data between the EDM and an aggregator. The gateway-to-aggregator link is serial and the gateway-to-EDM link uses SSH.

\textbf{Aggregator.} Each aggregator is an Arduino Nano 33 BLE. Aggregator-1 is the intended host of the TensorFlow Lite detector. The LM35 sensors have no wireless interface, so the sensor-to-aggregator links were wired in the experimental testbed.

\textbf{Sensors.} The sensors are LM35 precision temperature sensors, whose output voltage is proportional to temperature in degrees Celsius. Sensors attached to aggregator-1 were located indoors and provide the genuine data. Sensors attached to aggregator-2 were located outdoors, in a hotter environment, and provide the impostor-condition data.

\subsection{Threat Model and Assumptions}
Our security focus is the link between the sensors and their aggregator. The conceptual deployment assumes an unauthenticated wireless sensor-to-aggregator link. An attacker who joins that link can impersonate a genuine sensor by sending its own readings under the genuine sensor's identity. In the testbed, the impostor data reach aggregator-1 either through the EDM or directly from gateway-2, since both aggregators and the EDM share one local network. The detector intended for aggregator-1 flags incoming windows whose statistics deviate from the genuine sensor's learned behavior.

The gateway-to-EDM link runs over SSH. Breaking it would require the gateway's private key, which we leave outside the scope of the experiment. We also leave out an adversary who watches the genuine sensor and mimics its distribution, and physical attacks on the wired links.

\section{Detection Method}\label{sec:method}
\noindent
We treat detection as a prediction problem on the genuine sensor's time series. A model trained only on genuine readings predicts upcoming readings from a window of past ones. While the incoming stream comes from the genuine sensor the prediction error stays low. When it comes from a source that behaves differently, the error rises. A threshold on the mean absolute error (MAE) of each window turns this into a detection rule.

\subsection{Sequence Features}
The raw series is univariate. We slide a window over it and use each window of consecutive readings as the input and the reading that follows it as the target (Algorithm~\ref{alg:create_dataset}). We chose the window length by trial and error. The exact value is not preserved in the available records of the study.

\begin{algorithm}[!t]
\caption{Sliding-window feature extraction}
\label{alg:create_dataset}
\begin{algorithmic}[1]
\Require series $X$ of length $N$, window length $w$
\Ensure feature array $\mathbf{X}_s$, target array $\mathbf{y}_s$
\State $\mathbf{X}_s, \mathbf{y}_s \gets [\,], [\,]$
\For{$i \gets 0$ \textbf{to} $N - w - 1$}
    \State append $X[i : i+w]$ to $\mathbf{X}_s$
    \State append $X[i+w]$ to $\mathbf{y}_s$
\EndFor
\State reshape $\mathbf{X}_s$ to (samples, $w$, 1)
\State \Return $\mathbf{X}_s, \mathbf{y}_s$
\end{algorithmic}
\end{algorithm}

\subsection{Model}
The network has five layers, three LSTM layers followed by two fully connected layers. We built it in TensorFlow with the standard Keras LSTM layer, trained it with the MAE loss and the Adam optimizer, and used ReLU activation in the fully connected layers (Table~\ref{tab:hyp}). Layer widths and the learning rate were not recorded in the original source material.

For reference, an LSTM cell~\cite{hochreiter1997} with input $x_t$, previous hidden state $h_{t-1}$, and previous cell state $C_{t-1}$ computes
\begin{align}
f_t &= \sigma\big(W_f [h_{t-1}, x_t] + b_f\big), \label{eq:forget}\\
i_t &= \sigma\big(W_i [h_{t-1}, x_t] + b_i\big), \label{eq:input}\\
\tilde{C}_t &= \tanh\big(W_C [h_{t-1}, x_t] + b_C\big), \label{eq:cand}\\
C_t &= f_t \odot C_{t-1} + i_t \odot \tilde{C}_t, \label{eq:cell}\\
o_t &= \sigma\big(W_o [h_{t-1}, x_t] + b_o\big), \label{eq:output}\\
h_t &= o_t \odot \tanh(C_t), \label{eq:hidden}
\end{align}
where $\sigma$ is the logistic sigmoid and $\odot$ denotes elementwise multiplication. The forget gate $f_t$ scales the previous cell state, the input gate $i_t$ admits the candidate state $\tilde{C}_t$, and the output gate $o_t$ selects what part of the cell state is exposed as the hidden state. We chose LSTM layers over a plain recurrent layer for the cell state's ability to carry information across many steps.

\subsection{Threshold Rule}
Let $\mu_g$ and $s_g$ denote the mean and standard deviation of the per-window MAE obtained when the trained model is applied to genuine data. A window with MAE $e$ is flagged as impostor data when
\begin{equation}
e > \tau, \qquad \tau = \mu_g + 6\, s_g. \label{eq:threshold}
\end{equation}
We picked the factor of six for this data set after looking at the largest genuine MAE and the smallest impostor MAE (Section~\ref{sec:results}). The threshold was set once with the non-quantized TensorFlow Lite model and reused unchanged for the quantized variants.

\subsection{TensorFlow Lite Conversion and Quantization}
We converted the trained model with the TensorFlow Lite converter, which produces a FlatBuffer file that runs on the device without a round trip to a server. We produced three variants. The first keeps 32-bit floating-point weights. The second stores the weights as 16-bit floating-point values and restores them to 32-bit at load time. The third uses dynamic-range quantization and stores the weights as 8-bit integers. Weight quantization shrinks the model. What it does to latency and accuracy we measured rather than assumed (Section~\ref{sec:results}).

\section{Experimental Data}\label{sec:data}
\noindent
We first tried synthetic data generated from an ARMA model. It did not reproduce the behavior of the physical sensors, and all reported experiments use measured data from the testbed in Fig.~\ref{fig: Actual}.

\textbf{Genuine condition.} Readings were collected at a one-second interval from sensor $S_{11}$, attached to aggregator-1 indoors. A training series of 20,000 samples was collected first, followed by a validation series of 5,000 samples and a genuine test series of 3,600 samples. All three genuine sets come from the same indoor sensor.

\textbf{Impostor condition.} In parallel with the genuine test series, 3,600 samples were collected from sensor $S_{21}$, attached to aggregator-2 outdoors in a hotter environment.

\textbf{Interpretation.} The model sees genuine data only during training. The impostor series differs from the genuine series in two ways at once. It comes from a different device, and it was measured in a different thermal environment. The experiment does not separate these two effects, so the impostor condition amounts to a controlled distribution shift, and the results should be read as detection of that shift rather than as an evaluation of general device authentication. Fig.~\ref{fig:data} plots the four series and Table~\ref{tnab1} summarizes their statistics. The impostor series runs roughly ten degrees warmer than the genuine series on average, and the two ranges overlap only in their tails.

\section{Results}\label{sec:results}
\begin{figure}[!t]
    \centering
    \subfloat[Training data]{\includegraphics[width=0.49\columnwidth]{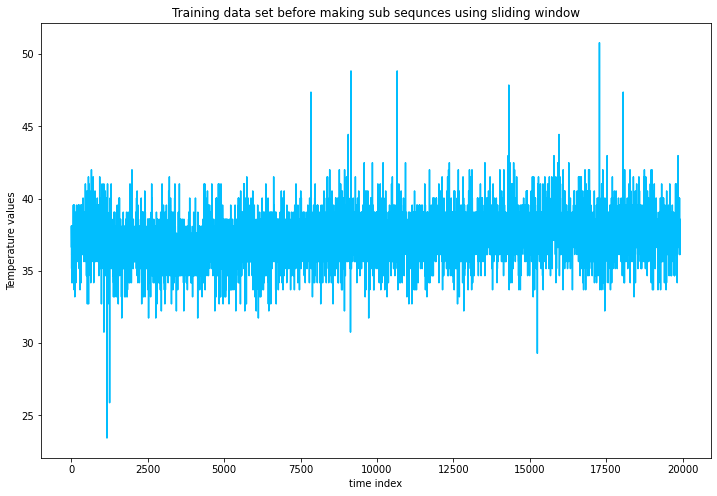}\label{nab1}}
    \hfill
    \subfloat[Validation data]{\includegraphics[width=0.49\columnwidth]{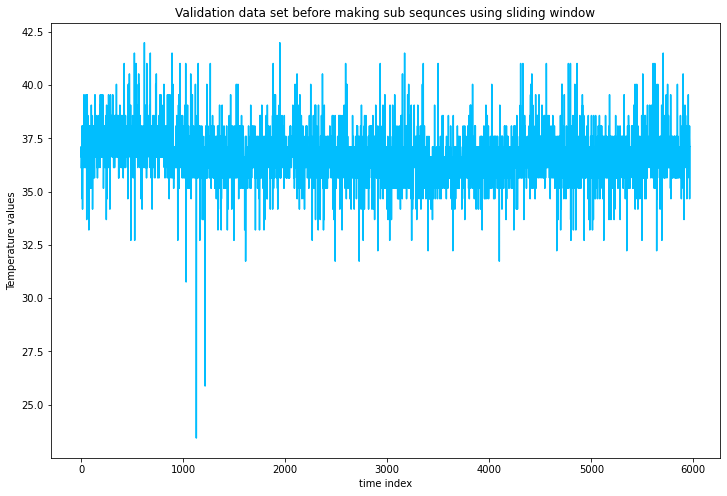}\label{nab2}}\\
    \subfloat[Genuine test data]{\includegraphics[width=0.49\columnwidth]{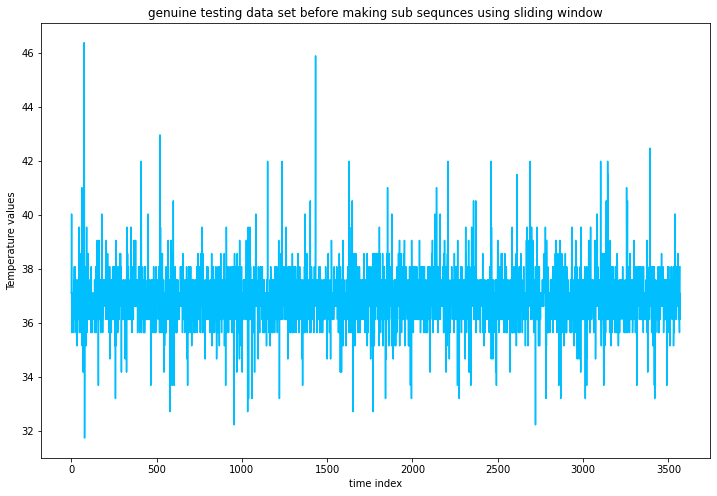}\label{nab3}}
    \hfill
    \subfloat[Impostor data]{\includegraphics[width=0.49\columnwidth]{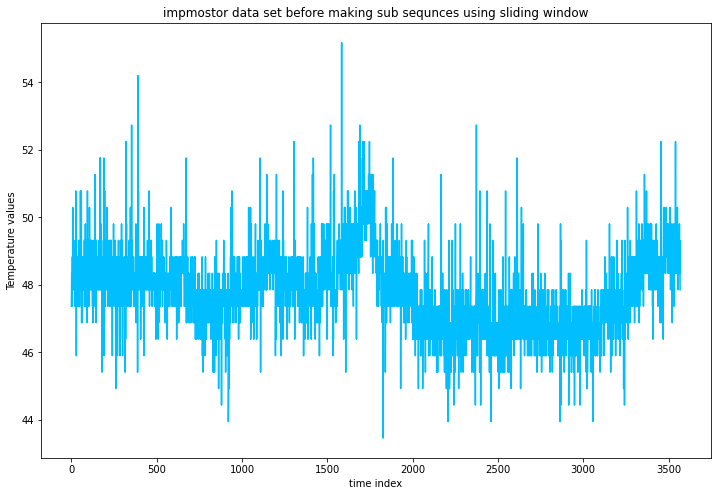}\label{nab4}}
    \caption{Raw temperature series before sliding-window feature extraction. Training, validation, and genuine test data come from the indoor sensor; impostor data come from the outdoor sensor.}
    \label{fig:data}
\end{figure}

\noindent
Fig.~\ref{fig:data} shows the training, validation, genuine test, and impostor series before windowing, and Table~\ref{tnab1} gives their summary statistics. Table~\ref{tab:hyp} lists the training settings. After training on the EDM, the model was converted to TensorFlow Lite in the three variants of Section~\ref{sec:method}. The three variants were evaluated with the TensorFlow Lite interpreter on the EDM. On-device deployment of the LSTM model to aggregator-1 was not completed with the tooling available at the time (Section~\ref{sec:discussion}).

\begin{table}[!t]
\centering
\caption{Summary statistics of the raw temperature series}
\label{tnab1}
\begin{tabular}{lcccc}
\toprule
Series & Min & Max & Mean & Std.\ dev. \\
\midrule
Training data      & 29.30 & 50.78 & 37.50 & 1.04 \\
Validation data    & 23.44 & 42.00 & 36.71 & 1.07 \\
Genuine test data  & 31.74 & 46.39 & 36.94 & 1.01 \\
Impostor test data & 43.46 & 55.18 & 47.80 & 1.20 \\
\bottomrule
\end{tabular}
\end{table}

\begin{table}[!t]
\centering
\caption{Training settings}
\label{tab:hyp}
\begin{tabular}{lc}
\toprule
Setting & Value \\
\midrule
Layers & 3 LSTM + 2 fully connected \\
Epochs & 150 \\
Batch size & 512 \\
Loss & MAE \\
Optimizer & Adam \\
Activation (fully connected) & ReLU \\
\bottomrule
\end{tabular}
\end{table}

We looked at each variant in three ways. Overlay plots show prediction quality on genuine and on impostor windows. A constructed transition sequence shows what happens when the source changes. Histograms show the distribution of per-window MAE on genuine and impostor data. The transition sequence has 1000 samples, the first 500 genuine test data and the last 500 impostor data, which puts the change point at index 500.

\subsection{Non-Quantized TensorFlow Lite Model}
\begin{figure*}[!t]
    \centering
    \subfloat[Genuine test data]{\includegraphics[width=0.42\textwidth]{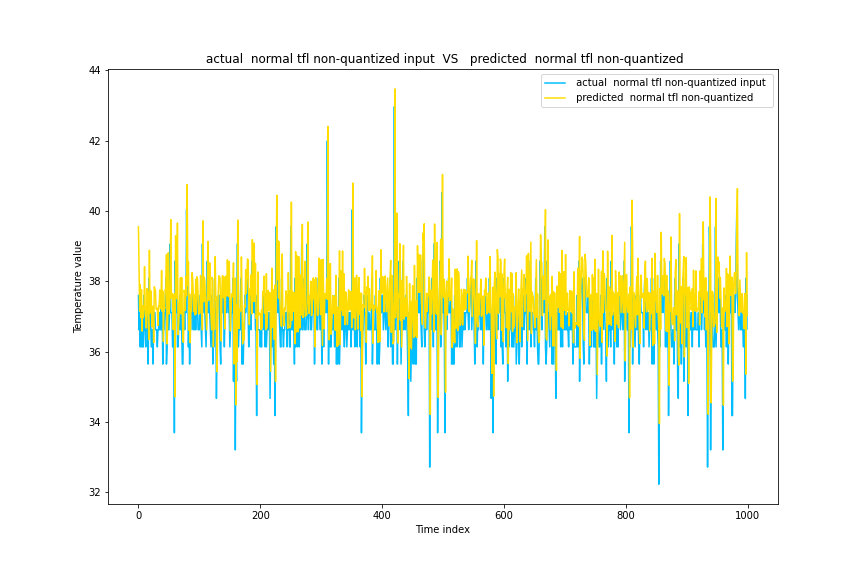}\label{nnq1}}
    \hfil
    \subfloat[Impostor data]{\includegraphics[width=0.42\textwidth]{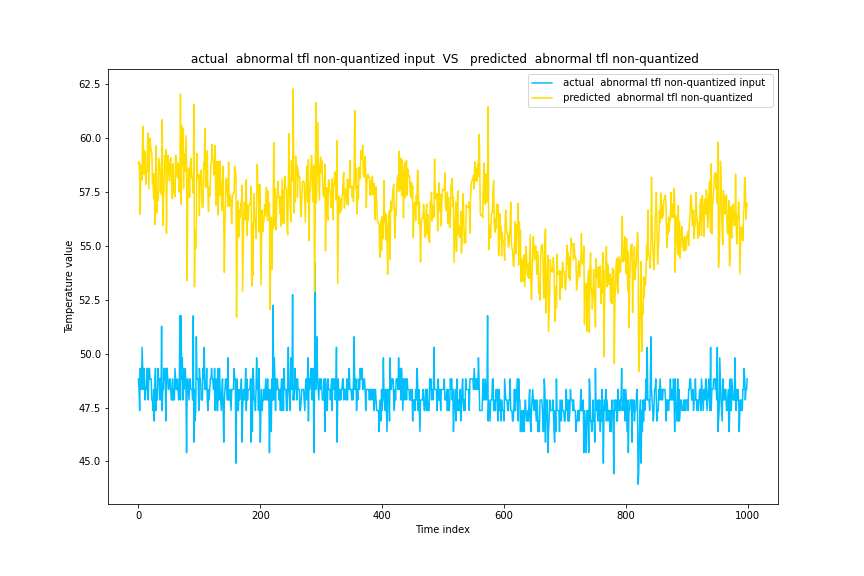}\label{innq1}}
    \caption{Non-quantized model: actual readings (blue) and model predictions (yellow).}
    \label{lnq1}
\end{figure*}

\begin{figure*}[!t]
    \centering
    \subfloat[Non-quantized]{\includegraphics[width=0.32\textwidth]{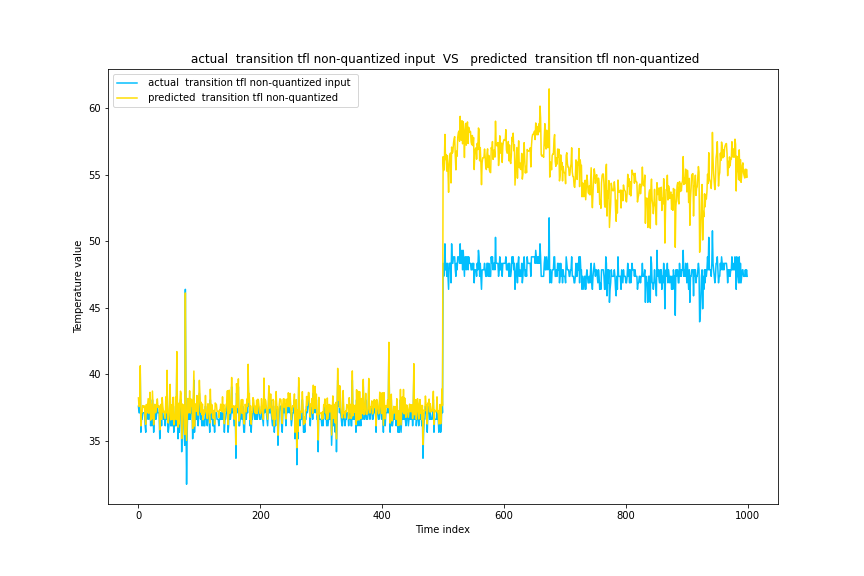}\label{lnq2}}
    \hfil
    \subfloat[16-bit weights]{\includegraphics[width=0.32\textwidth]{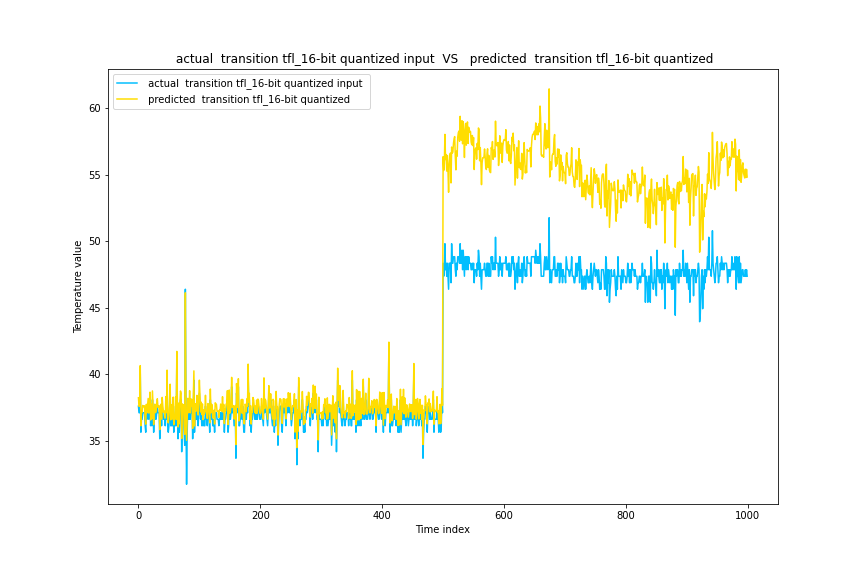}\label{lq162}}
    \hfil
    \subfloat[8-bit weights]{\includegraphics[width=0.32\textwidth]{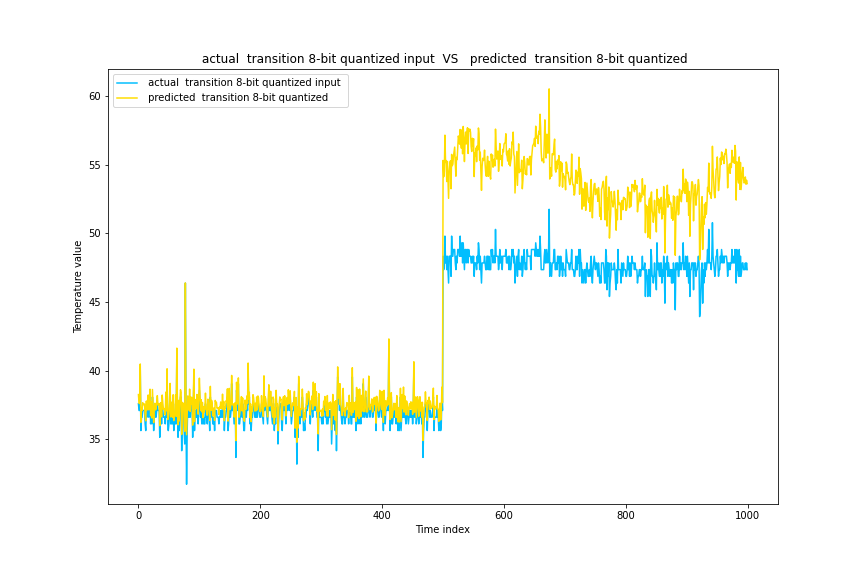}\label{lq82}}
    \caption{Transition sequence for the three variants: genuine data for indices 0--499, impostor data for indices 500--999. Actual readings in blue, predictions in yellow. In all three cases prediction error rises at the change point.}
    \label{fig:transitions}
\end{figure*}

Fig.~\ref{lnq1} overlays actual and predicted values. On genuine data the two series track each other closely. On impostor data the prediction sits several degrees away from the actual series throughout. On the transition sequence (Fig.~\ref{fig:transitions}(a)) the prediction error is small before index 500 and jumps right after it, so the model flags the transition at the constructed change point.

\begin{figure*}[!t]
    \centering
    \subfloat[Genuine test windows]{\includegraphics[width=0.42\textwidth]{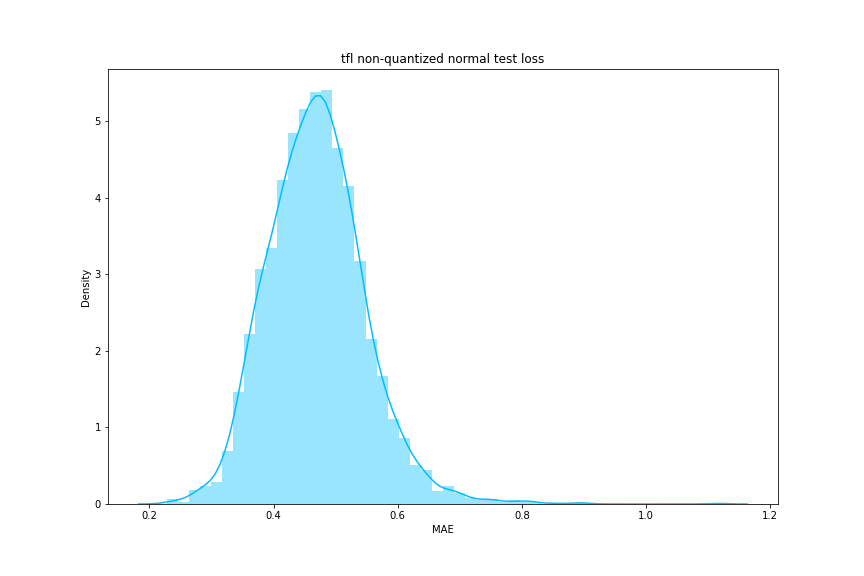}\label{nnq3}}
    \hfil
    \subfloat[Impostor windows]{\includegraphics[width=0.42\textwidth]{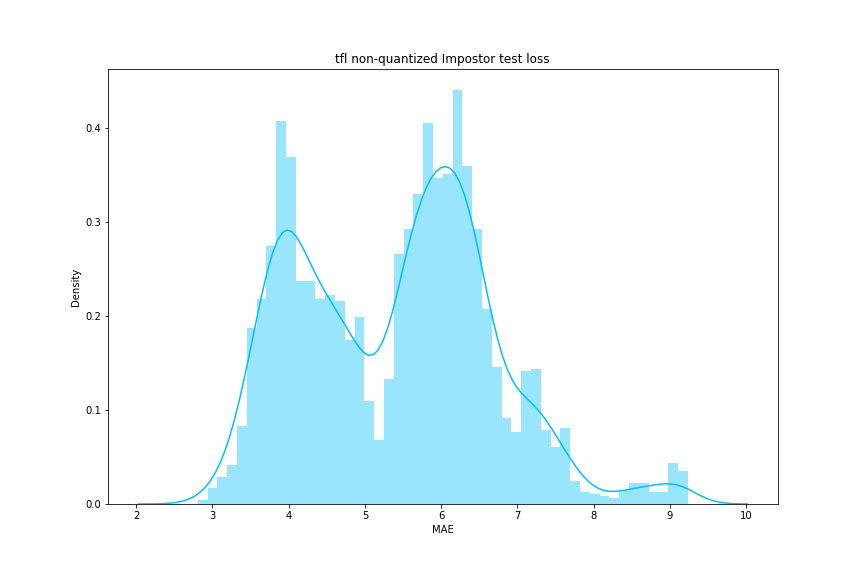}\label{inq3}}
    \caption{Non-quantized model: histogram and density of per-window MAE. Note the different horizontal scales.}
    \label{lnnq3}
\end{figure*}

Fig.~\ref{lnnq3} shows histograms of the per-window MAE. Genuine windows cluster at small MAE values and impostor windows at values several times larger, with a visible gap between the two. We chose the threshold in~\eqref{eq:threshold} by comparing the largest genuine MAE with the smallest impostor MAE. Six standard deviations above the genuine mean landed in this gap for this data set. The same threshold is reused for the quantized variants below.

\subsection{16-Bit and 8-Bit Weight-Quantized Models}
\begin{figure*}[!t]
    \centering
    \subfloat[16-bit, genuine test data]{\includegraphics[width=0.42\textwidth]{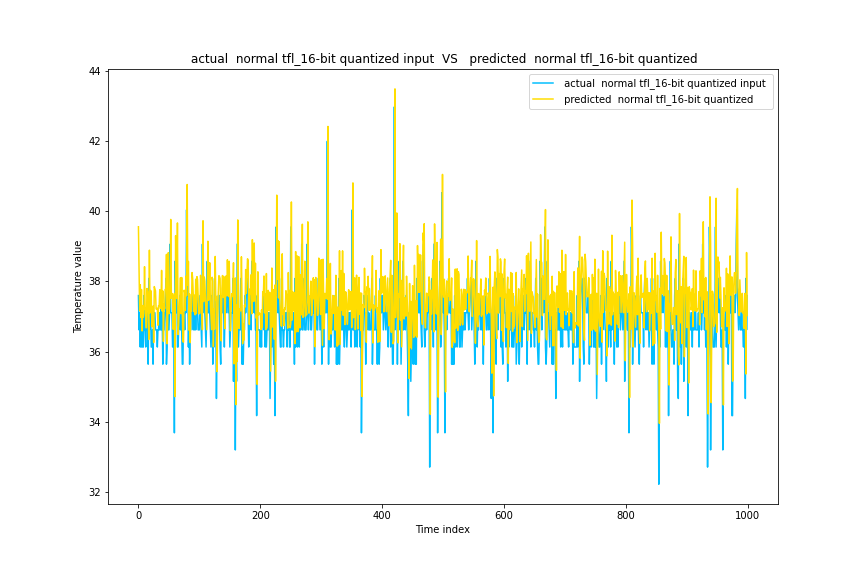}\label{nq161}}
    \hfil
    \subfloat[16-bit, impostor data]{\includegraphics[width=0.42\textwidth]{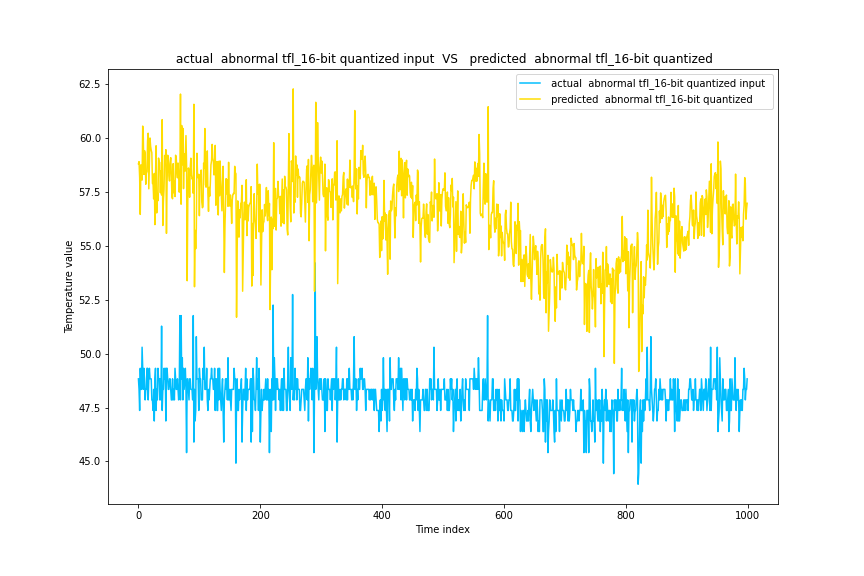}\label{inq161}}\\
    \subfloat[8-bit, genuine test data]{\includegraphics[width=0.42\textwidth]{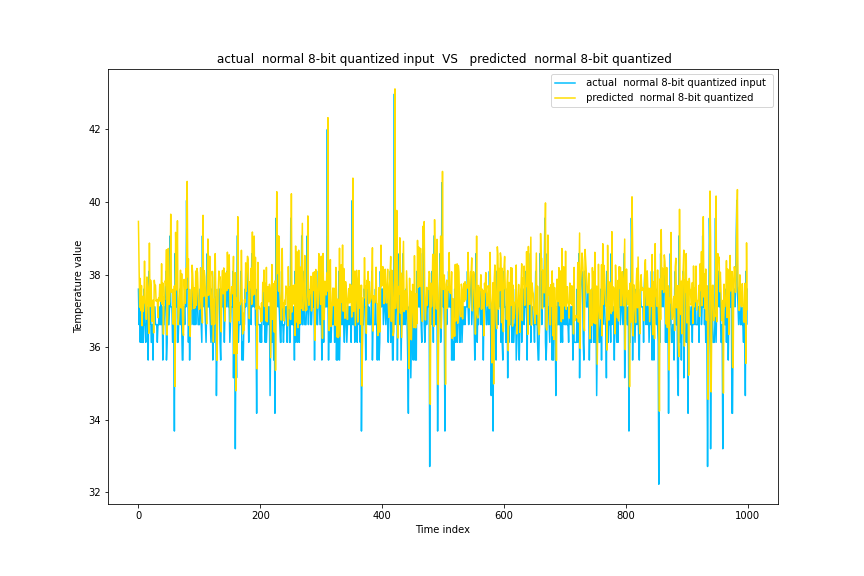}\label{nq81}}
    \hfil
    \subfloat[8-bit, impostor data]{\includegraphics[width=0.42\textwidth]{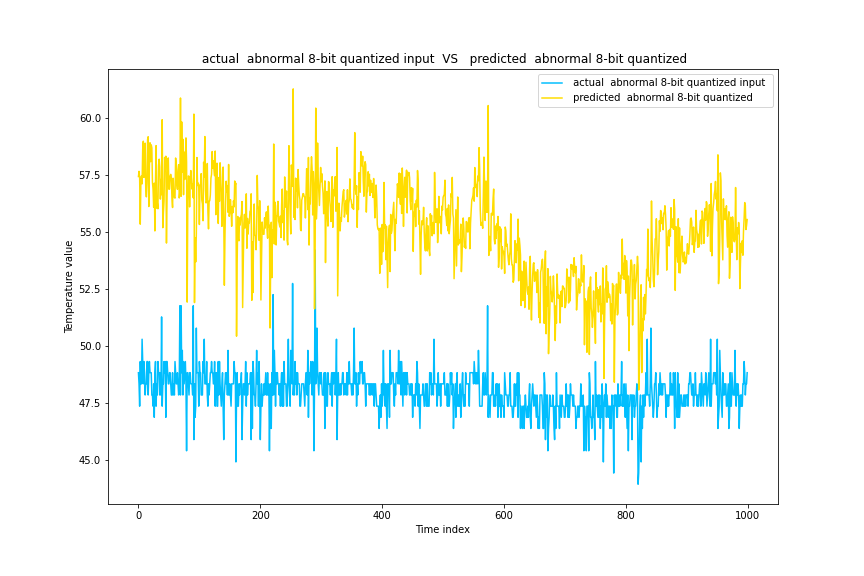}\label{inq81}}
    \caption{Weight-quantized models: actual readings (blue) and predictions (yellow).}
    \label{fig:qoverlay}
\end{figure*}

\begin{figure*}[!t]
    \centering
    \subfloat[16-bit, genuine test windows]{\includegraphics[width=0.42\textwidth]{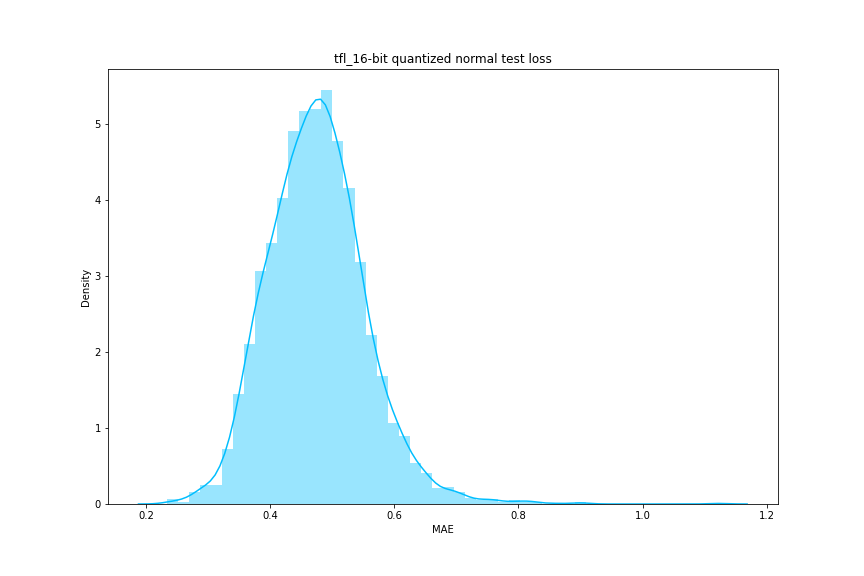}\label{nq163}}
    \hfil
    \subfloat[16-bit, impostor windows]{\includegraphics[width=0.42\textwidth]{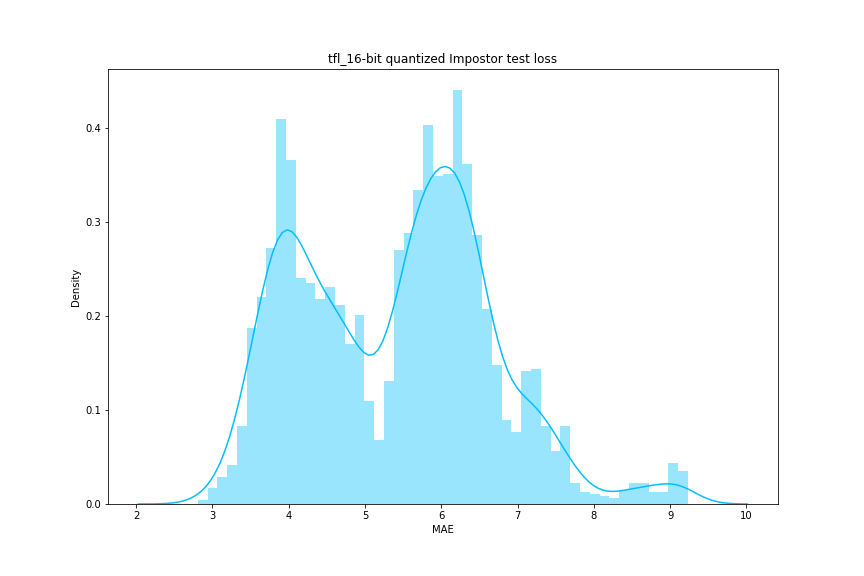}\label{iq163}}\\
    \subfloat[8-bit, genuine test windows]{\includegraphics[width=0.42\textwidth]{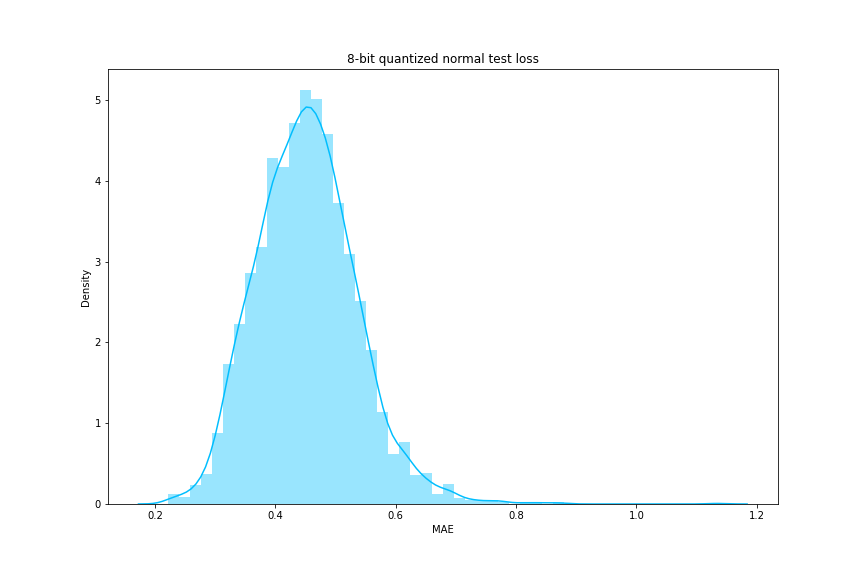}\label{nq83}}
    \hfil
    \subfloat[8-bit, impostor windows]{\includegraphics[width=0.42\textwidth]{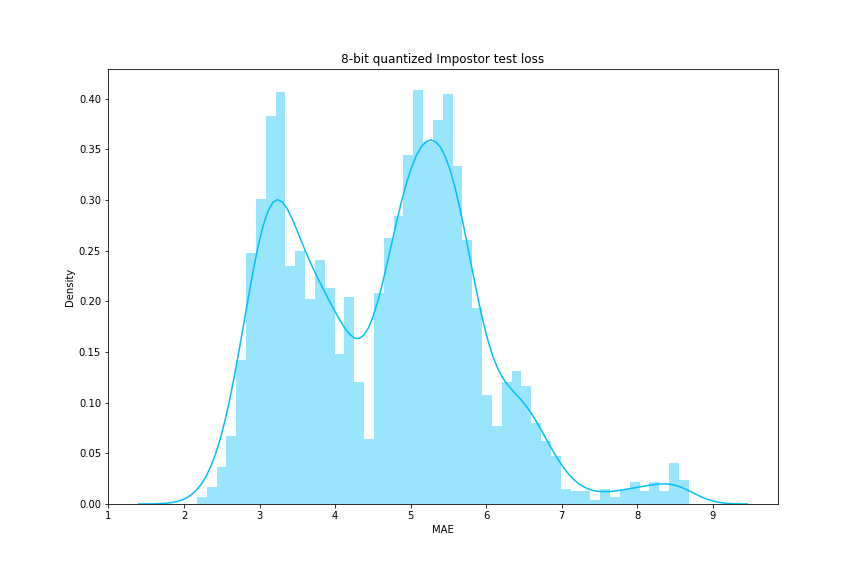}\label{iq83}}
    \caption{Weight-quantized models: histogram and density of per-window MAE.}
    \label{fig:qmae}
\end{figure*}

With 16-bit weights, the overlay plots (Fig.~\ref{fig:qoverlay}(a,b)), the transition sequence (Fig.~\ref{fig:transitions}(b)), and the MAE histograms (Fig.~\ref{fig:qmae}(a,b)) look almost the same as for the non-quantized model, and the transition is again flagged at index 500. The 8-bit model (Figs.~\ref{fig:qoverlay}(c,d), \ref{fig:transitions}(c), and \ref{fig:qmae}(c,d)) also looks much the same and flags the transition at index 500. The numbers in Table~\ref{tab:nqopim} tell a slightly different story, with a larger accuracy change for the 8-bit variant than for the 16-bit one.

\subsection{Performance Comparison}
Table~\ref{tab:nqopim} summarizes size, inference time, and detection accuracy for the three variants. The original TensorFlow model took 1.53 MB. Converting it to TensorFlow Lite alone brought it down to 554.06 KB, about 64.7\% smaller. Storing the weights as 16-bit floating point cut it further to 298.5 KB, and 8-bit dynamic-range quantization to 185.08 KB, about 11.8\% of the original TensorFlow model. Detection accuracy was 99.980\% for the non-quantized model, 99.972\% for the 16-bit model, and 98.206\% for the 8-bit model.

Inference time went the other way. Processing the 3569 test windows took 11.296 s with the non-quantized model, 15.226 s with the 16-bit model, and 18.544 s with the 8-bit model. In our experiment, weight quantization traded model size for inference time rather than reducing both. We did not isolate the cause of the timing difference.

The accuracy values are the ones recorded in the original experiment. We obtained them by applying the threshold rule~\eqref{eq:threshold} to the test windows and counting the windows that fell on the correct side of the threshold. The original records do not say whether this count covered the impostor windows alone or genuine and impostor windows together. The inference timings were measured with the TensorFlow Lite interpreter on the EDM, not on the microcontroller.

\begin{table}[!t]
\centering
\caption{Size, inference time, and detection accuracy of the three TensorFlow Lite variants}
\label{tab:nqopim}
\begin{tabular}{lccc}
\toprule
Metric & Non-quantized & 16-bit & 8-bit \\
\midrule
Model size (KB) & 554.06 & 298.5 & 185.08 \\
Inference time, 3569 windows (s) & 11.296 & 15.226 & 18.544 \\
Detection accuracy (\%) & 99.980 & 99.972 & 98.206 \\
\bottomrule
\end{tabular}
\end{table}

\section{Discussion}\label{sec:discussion}
\noindent
The overlay plots in Figs.~\ref{lnq1} and \ref{fig:qoverlay} show that a model trained on genuine data only tracks genuine readings closely and misses impostor readings by several degrees. The gap between genuine and impostor windows in the MAE histograms (Figs.~\ref{lnnq3} and \ref{fig:qmae}) is what makes a simple threshold work here. How well the rule does comes down to the overlap between the two MAE distributions. With as little overlap as in this data set, it separates the two conditions almost perfectly. The transition sequences (Fig.~\ref{fig:transitions}) show all three variants flagging the change at the constructed change point.

These results need to be read with the data collection in mind. The impostor sensor sat outdoors in a hotter environment, and its readings average about ten degrees warmer than the genuine ones (Table~\ref{tnab1}). That alone pushes the two MAE distributions far apart, and the high accuracies in Table~\ref{tab:nqopim} reflect this separation. What the experiment shows is that the deployed model detects a pronounced, controlled distribution shift. It does not show how the detector would fare against an impostor whose readings resemble the genuine sensor's, or against an adversary who adapts to the detector.

Table~\ref{tab:nqopim} also shows that quantization was not free for us. It cut model size substantially, with a small accuracy change for 16-bit weights and a larger one for 8-bit weights, but measured inference time went up for both quantized variants. We did not isolate the cause of the timing difference.

During development we tried convolutional layers on the windowed data. They did not give good predictions, and we settled on the LSTM-based model instead. Getting the LSTM model onto the microcontroller was the step that did not complete. TensorFlow Lite for Microcontrollers had no kernel for the fused LSTM operation when the work was done, and the model's tensor arena requirements were a further obstacle on a board with 256 KB of SRAM. Whether this model fits within the memory of the Nano 33 BLE was not re-evaluated for this release.

\subsection{Limitations}
Several limitations bound what can be concluded from this study.
\begin{itemize}
    \item \textbf{One sensor modality and a univariate series.} Only temperature from LM35 sensors was used, and the model consumes a single scalar series.
    \item \textbf{Controlled indoor/outdoor shift.} The genuine and impostor sensors were in different thermal environments.
    \item \textbf{Rule-based threshold selected on the test data.} The threshold of six standard deviations above the mean genuine MAE was selected after inspecting the genuine and impostor MAE distributions used in this experiment rather than on an independent calibration set. The reported accuracy should therefore be interpreted as descriptive performance on this controlled data set rather than as an unbiased estimate of out-of-sample detection performance.
    \item \textbf{Small testbed.} One genuine sensor and one impostor sensor, one aggregator as the intended detector host, and one collection period were used. Results are from a single training run.
    \item \textbf{No adaptive adversary.} The threat model assumes an impostor that simply injects its own readings. Attackers who observe and mimic the genuine sensor were not evaluated.
   
\end{itemize}

\subsection{Future Work}
The directions we identified at the time were multivariate sensor data, automatic threshold selection, online learning and model updates on the device, tests with impostor sensors placed in the same environment as the genuine sensor, and tests against adversaries that try to mimic genuine behavior.

\section{Conclusion}\label{sec:conclusion}
\noindent
We have described a proof-of-concept study of sensor impersonation detection at the edge. We formulate detection as sequence prediction on a genuine sensor's time series. A model with three LSTM layers and two fully connected layers, trained on genuine temperature readings only, flags a window when its prediction MAE exceeds the mean genuine MAE by more than six standard deviations. We converted the model to TensorFlow Lite for an Arduino Nano 33 BLE aggregator in non-quantized, 16-bit, and 8-bit weight-quantized variants and evaluated the variants with the TensorFlow Lite interpreter on the local server. On a controlled testbed with the impostor sensor in a hotter outdoor location, the three variants reached detection accuracies of 99.980\%, 99.972\%, and 98.206\% and flagged the change point of a constructed genuine-to-impostor sequence. Quantization shrank the model from 554 KB to 298.5 KB and 185 KB but made measured inference slower.

The experiment shows that a small LSTM-based predictor with a simple threshold can be converted to a microcontroller-sized TensorFlow Lite model and can pick up a pronounced distribution shift in a sensor stream. Executing the LSTM model on the microcontroller itself was not achieved with the tooling available at the time. The genuine and impostor conditions were well separated, so the results say nothing about general device authentication or about adversaries that mimic genuine behavior. Impostors in the same environment, multivariate data, and automatic threshold selection are left for future work.

\section*{Acknowledgment}
The author thanks Prof. Michael Perkins, who advised this research at Carnegie Mellon University, and Prof. Timothy Brown, and acknowledges support from the Mastercard Foundation.

\bibliographystyle{IEEEtran}
\bibliography{main}

\end{document}